\documentclass[usenatbib]{mn2e}
\usepackage{rotating,pdflscape}
\usepackage{graphicx}
\usepackage{amssymb,amsmath}
\usepackage{natbib}
\begin{document}

\title[LVL SEDs and Physical Properties]{Spitzer Local Volume Legacy (LVL) SEDs and Physical Properties}

\author[D. Cook et al.]
{David O. Cook,$^1$
Daniel A. Dale,$^1$
Benjamin D. Johnson,$^2$
Liese Van Zee,$^3$
\newauthor
Janice C. Lee,$^4$
Robert C. Kennicutt,$^{5,6}$
Daniela Calzetti,$^7$
Shawn M. Staudaher$^1$
\newauthor
 and Charles W. Engelbracht$^{8}$\\
$^1$Department of Physics \& Astronomy, University of Wyoming, Laramie, WY 82071, USA; dcook12$@$uwyo.edu\\
$^2$UPMC-CNRS, UMR7095, Institut d'Astrophysique de Paris, F-75014, Paris, France\\
$^3$Department of Astronomy, Indiana University, Bloomington, IN 47405, USA\\
$^4$Space Telescope Science Institute, 3700 San Martin Drive, Baltimore, MD 21218, USA\\
$^5$Institute of Astronomy, University of Cambridge, Cambridge CB3 0HA, UK\\
$^6$Steward Observatory, University of Arizona, Tucson, AZ 85721, USA\\
$^7$Department of Astronomy, University of Massachusetts, Amherst, MA 01003, USA\\
$^8$Raytheon Company, 1151 East Hermans Road, Tucson, AZ 85756, USA; Deceased}

\maketitle
\begin{abstract}
We present the panchromatic spectral energy distributions (SEDs) of the Local Volume Legacy (LVL) survey which consists of 258 nearby galaxies ($D<$11~Mpc). The wavelength coverage spans the ultraviolet to the infrared (1500~$\textrm{\AA}$ to 24~$\mu$m) which is utilized to derive global physical properties (i.e., star formation rate, stellar mass, internal extinction due to dust.). With these data, we find color-color relationships and correlated trends between observed and physical properties (i.e., optical magnitudes and dust properties, optical color and specific star formation rate, and ultraviolet-infrared color and metallicity). The SEDs are binned by different galaxy properties to reveal how each property affects the observed shape of these SEDs. In addition, due to the volume-limited nature of LVL, we utilize the dwarf-dominated galaxy sample to test star formation relationships established with higher-mass galaxy samples. We find good agreement with the star-forming ``main-sequence" relationship, but find a systematic deviation in the infrared ``main-sequence" at low luminosities. This deviation is attributed to suppressed polycyclic aromatic hydrocarbon (PAH) formation in low metallicity environments and/or the destruction of PAHs in more intense radiation fields occurring near a suggested threshold in sSFR at a value of log($sSFR$) $\sim$ --10.2. 
\end{abstract}


\section{Introduction}
Low-mass galaxies are excellent laboratories in which extreme star formation and dust environments can be studied. These galaxies have some of the lowest star formation rates (SFRs), lowest metallicities, and highest specific star formation rates (sSFR) \citep[e.g.,][]{kennicutt08,marble10,lee11,weisz11,berg12}. Comparing the star formation relationships of low-mass galaxy samples to those with higher-mass galaxies can constrain the underlying physics of the star formation process.

The Local Volume Legacy (LVL) survey has provided large data sets which include ultraviolet \citep[UV;][]{lee11} non-ionizing stellar continuum, H$\alpha$ nebular emission \citep{kennicutt08}, and infrared \citep[IR;][]{dale09} dust emission of hundreds of local galaxies ($D<11$~Mpc). Due to the volume-limited nature of LVL, the galaxy sample consists of large spirals and a few elliptical galaxies, but is dominated by low-mass dwarf and irregular galaxies where 52\% of the sample has a stellar mass less than $10^9 $M$_{\odot}$ \citep{dale09}. As a result of the LVL sample composition, any star formation or dust relationships found in the extreme, low-mass environments can be directly compared to larger LVL galaxies with the same methodology. 


The ``main sequence" of star-forming galaxies shows a tight correlation between stellar mass and SFR \citep[e.g.,][]{brinchmann04,daddi07,elbaz07,noeske07,salim07,peng10,klee12,leitner12,heinis14}. The majority of star-forming galaxies reside on this sequence; however, no previous study has probed down to the lowest stellar mass measured by the LVL sample ($M_{\star}\sim 10^{6}~$M$_{\odot}$). Since low-mass galaxies have extreme environments (e.g., low SFRs, low metallicities, etc.), it is important to examine if they follow the same ``main-sequence" relationship.

The ``main sequence" is also represented as a ``blue cloud" of galaxies found in the optical color-magnitude (similarily the color-$M_{\star}$) diagram. Many studies have found a bimodality in these diagrams where optically blue (or ``blue cloud") galaxies represent the ``main sequence" with active star formation, optically red (or ``red sequence") galaxies represent a population with highly reduced star formation activity, and intermediate optical color (or ``green-valley") galaxies represent a population where star formation is in the process of being quenched \citep[e.g.,][]{bell04,faber07,martin07,salim07,schiminovich07,wyder07,Goncalves12,schawinski14}. The SFR trend from the ``blue cloud" through the ``green valley" to the ``red sequence" implies an evolutionary process regulated by SFR. However, some studies have found that low-mass galaxies can have circuitous paths across the color-magnitude diagram \citep{eskew11}.
 
Furthermore, \citet{schawinski14} found that SDSS ``green-valley" galaxies identified by their optical colors are located significantly below the ``main-sequence" on the SFR versus $M_{\star}$ diagram indicating a reduced SFR for a given stellar mass. Since all previous studies have investigated this evolution in higher-mass galaxy samples compared to ours, it is important to examine if low-mass galaxies exhibit similar evolutionary trends via the existence of low-mass, ``green-valley" galaxies.

It is also important to examine how galaxies evolve over cosmic time (i.e., at higher redshifts). Studies of higher redshift galaxies have shown increased SFRs at a given stellar mass (see Figure~\ref{fig:MstarSFR}) indicating that these galaxies are scaled-up versions of local galaxies \citep[e.g.,][]{elbaz07,noeske07,heinis14}. Due to the increasing number of galaxy mergers back in cosmic time to a redshift of $\sim$1$-$2, an increasing fraction of higher redshift galaxies show increased SFRs and IR luminosities referred to as luminous and ultra luminous infrared galaxies \citep[LIRGs and ULIRGs, respectively;][]{chary01,lefloch05,magnelli09}. Furthermore, many studies \citep[][]{daddi07,papovich07,magnelli11} have found that these galaxies show an excess of mid-IR luminosities (attributed to an increase of 8$\mu$m emission from PAH molecules) compared to the total-IR (TIR; integrated 8 to 1000$\mu$m) luminosities. These findings suggest that the IR SEDs of higher redshift galaxies exhibit different fundamental IR properties and that their IR SED shapes evolve over cosmic time.

However, \citet{elbaz11} used longer wavelength observations which more accurately describes the peak of the IR SED to normalize the mid-IR luminosities (i.e., 8$\mu$m fluxes) and found that the ratio forms a constant ``main sequence" of IR galaxies which is regulated by SFR. The redshift range ($0 < z < 2.5$) and luminosity range ($9 < $ log$(TIR) < 12~$L$_{\odot}$) probed by \citet{elbaz11} suggests that the mid-IR excess scales with the total infrared luminosity for most star-forming galaxies. It should be noted that the majority of the \citet{elbaz11} sample consists of (U)LIRGs with a TIR luminosity greater than $10^{10}$ L$_{\odot}$. If the ``IR main sequence" is an analog to the ``main sequence" of star-forming galaxies, then testing its universality with low-mass galaxies provides an important constraint. 

This paper constructs the panchromatic (UV-optical-IR) SEDs which facilitates examination of relationships between different observable and physical properties. In addition, we utilize the low-luminosity ($5.1 < $ log$(TIR) < 10.8~$L$_{\odot}$) dominated LVL sample to test star formation relationships established with more luminous galaxy samples. Finally, deviations in one of these relationships lead to an investigation of how the extreme environments of low-mass galaxies affect their mid-IR polycyclic aromatic hydrocarbon (PAH) emission. 

\section{Sample \label{sec:sample}}
The LVL sample consists of 258 of our nearest galaxy neighbors reflecting a statistically complete, representative sample of the local Universe. The sample selection and description are detailed in \citet{dale09}, but we provide an overview here. 

The LVL sample was built upon the samples of two previous nearby galaxy surveys: ACS Nearby Galaxy Survey Treasury \citep[ANGST;][]{dalcanton09} and 11~Mpc H$\alpha$ and Ultraviolet Galaxy Survey \citep[11 HUGS;][]{kennicutt08,lee11}. The final LVL sample consists of galaxies that appear outside the Galactic plane ($|b| > 20^{\circ}$), have a distance within 11~Mpc ($D \leq$ 11~Mpc), span an absolute $B-$band magnitude of $-9.6 < M_B < -20.7$ mag, and span an RC3 catalog galaxy type range of $-5 < T < 10$. Although the galaxies span a wide range in galaxy type, the sample is dominated by dwarf galaxies due to its volume-limited nature. The full LVL sample and basic properties are listed in Table~\ref{tab:genprop}.

\begin{table*}
{General Galaxy Properties}
\begin{tabular}{lcccccccc}
\hline
\hline
Galaxy & RA        & DEC        & D     & T   & 12+log(O/H) & Method & $A_{\rm{FUV}}$ & log($M_{\star}$) \\
Name   & (J2000.0) & (J2000.0)  & (Mpc) &     &             &        & (mag)          & (M$_{\odot}$)    \\
(1)    & (2)       & (3)        & (4)   & (5) & (6)         & (7)    & (8)            & (9)              \\
\hline
                 WLM   & 00:01:58.16   &   -15:27:39.3   &   ~0.92   &     ~10   &                        7.8$\pm$0.10   &       4363   &                               0.055   &    ~7.42  \\ 
             NGC0024   & 00:09:56.54   &   -24:57:47.3   &   ~8.13   &     ~~5   &     \hspace{0.15cm}8.9$\pm$0.11$^a$   &     Strong   &                               0.325   &    ~9.37  \\ 
             NGC0045   & 00:14:03.99   &   -23:10:55.5   &   ~7.07   &     ~~8   &                              \ldots   &     \ldots   &                               0.139   &    ~9.55  \\ 
             NGC0055   & 00:14:53.60   &   -39:11:47.9   &   ~2.17   &     ~~9   &                        8.4$\pm$0.10   &     Strong   &                               0.435   &    ~9.53  \\ 
             NGC0059   & 00:15:25.13   &   -21:26:39.8   &   ~5.30   &    $-$3   &                        8.4$\pm$0.11   &       4363   &                               0.632   &    ~8.52  \\ 
         ESO410-G005   & 00:15:31.56   &   -32:10:47.8   &   ~1.90   &    $-$1   &                              \ldots   &     \ldots   &            \hspace{0.15cm}0.336$^c$   &    ~7.14  \\ 
        SCULPTOR-DE1   & 00:23:51.70   &   -24:42:18.0   &   ~4.20   &     ~10   &                              \ldots   &     \ldots   &            \hspace{0.15cm}0.336$^c$   &   \ldots  \\ 
         ESO294-G010   & 00:26:33.37   &   -41:51:19.1   &   ~1.90   &    $-$3   &                              \ldots   &     \ldots   &            \hspace{0.15cm}0.336$^c$   &    ~6.79  \\ 
              IC1574   & 00:43:03.82   &   -22:14:48.8   &   ~4.92   &     ~10   &                              \ldots   &     \ldots   &            \hspace{0.15cm}0.336$^c$   &    ~7.70  \\ 
             NGC0247   & 00:47:08.55   &   -20:45:37.4   &   ~3.65   &     ~~7   &                              \ldots   &     \ldots   &                               0.232   &    ~9.60  \\ 
\hline
\end{tabular} \\
\caption[]{Column 1: Galaxy name. Column 2 and 3: J2000 right ascension and declination as tabulated by the NASA/IPAC Extragalactic Database$^{\dag}$ in 2014 July. Column 4: distance in Mpc from \citet{kennicutt08}. Column 5: RC3 Morphological T-type from \citet{kennicutt08}. Column 6: Oxygen abundances drawn from the literature: $^{\rm{a}}$\citet{moustakas10}, $^{\rm{b}}$\citet{berg12}, and no superscript represent values from \citet{marble10}. Column 6: Method of oxygen abundance determination which are based on either the 4363~$\rm{\AA}$ line, the strong-line method, a mix of these two methods, or, in one case, a planetary nebula observation. Column 7: Extinction due to dust in the GALEX FUV bandpass presented in units of magnitudes where $^{\rm{c}}$ represents extinction corrections dervied from $A_{\rm{H\alpha}}$. Column 8: The stellar mass in logarithmic solar units derived from the 3.6$\mu m$ luminosity. The full table is available online.

$^{\dag}$ http://ned.ipac.caltech.edu/

}

\label{tab:genprop}
\end{table*}

\section{Data \label{sec:data}}
The LVL data consist of imaging from GALEX ultraviolet \citep[UV;][]{lee11}, ground-based H$\alpha$ \citep{kennicutt08}, ground-based optical \citet{cook14a}, 2MASS near-infrared \citep[NIR;][]{dale09}, and Spitzer infrared \citep[IR;][]{dale09} observations. We use these data to construct the panchromatic ($FUV-24\mu m$) galaxy SEDs where the magnitudes in each bandpass are in the AB magnitude system and have been corrected for Milky Way foreground extinction using the \citet{schlegel98} maps and the dust reddening curve of \citet{draine03}.

There are three previously published galaxy apertures defined at UV, optical, and IR wavelengths. The optical apertures are taken from the RC3 catalog \citep{rc3} as tabulated by the VizieR catalog,\footnote{http://vizier.u-strasbg.fr/} and are defined as isophotal ellipses with surface brightness of 25 mag/arcsec$^2$ in the $B-$band filter. The optical apertures are tabulated in \citet{cook14a}. 

The Spitzer Space Telescope IR elliptical apertures of \citep{dale09} were chosen to encompass the majority of the emission seen at GALEX UV (1500\AA--2300\AA) and Spitzer IR (3.6$\mu$m--160$\mu$m) wavelengths. In practice, these apertures were usually determined by the extent of the 3.6$\mu$m emission given the superior sensitivity of the 3.6$\mu$m array coupled with the relatively bright emission from older stellar populations at this wavelength. However, in several instances the emission between 1500\AA--2300\AA~or at 160$\mu$m wavelengths were spatially more extended, and thus these wavelengths were used to determine the IR apertures. The resulting median ratio of IR-to-optical semi-major axes is 1.5.

The GALEX UV elliptical apertures of \citet{lee11} were defined as an isophotal ellipse outside of which the photometric error was greater than 0.8 mag or the intensity fell below the sky level. The resulting median ratio of UV-to-optical semi-major axes is 2.3. 

To make direct photometric comparisons across multiple wavelengths (UV--optical--IR) we have chosen to present figures based on the fluxes within one aperture: the IR apertures of \citet{dale09}. These apertures are chosen for the following reasons: (1) the IR apertures yields similar global fluxes to those of the UV apertures, both of which exhibit greater fluxes compared to the fluxes measured within the R25 apertures \citet{cook14a}; (2) there are fewer galaxy apertures ($N=234$) defined in the UV catalog of \citet{lee11} compared to the number of IR apertures ($N=255$) published in \citet{dale09}; (3) \citet{lee11} measured the global UV fluxes within both the UV and IR apertures while the fluxes within the UV apertures were not measured by \citet{dale09} since the UV apertures were published at a later date. The IR apertures have been visually checked to ensure that they are adequately large to encompass all of the optical and typically all of the UV and IR emission. 

There are 1, 13, and 47 upper limit data points published in the optical \citet{cook14a}, UV \citep{lee11}, and IR \citep{dale09} studies, respectively. No upper limits are shown in the figures of \S\ref{sec:results} and \S\ref{sec:discussion} where the majority of these figures present flux ratios or colors of various fluxes. Galaxies with non-detections in both fluxes of these ratios and colors do not provide any constraints and galaxies with a non detection in only one measurement are not shown to more clearly present our results. However, the few upper limit data points with only a single non detection either agree with or are consistent with the trends presented in \S\ref{sec:results} and \S\ref{sec:discussion}.

\section{Physical Properties \label{sec:physprop}}
Here we present the physical properties of the LVL galaxy sample obtained from previous LVL publications (e.g., metallicity) in addition to properties derived in this publication (e.g., internal dust extinction, SFR, $M_{\star}$, and stellar continuum subtracted PAH emission).

\subsection{Internal Dust Extinction \label{sec:ext}}
To make comparisons that utilize photometry which are unaffected by the attenuation due to dust, we present analyses in this paper where the UV-optical-IR fluxes are corrected for internal extinction. These corrections are carried out via the prescription of \citet{hao11} where the extinction in the FUV bandpass ($A_{\rm{FUV}}$) is first calculated via the empirical relationship with the 24$\mu m$-to-FUV luminosity ratio ($L_{24}$/$L_{FUV}$). All other bandpass extinctions ($A_{\rm{\lambda}}$) are derived from $A_{\rm{FUV}}$ in combination with the dust extinction curve of \citet{draine03}. 

Not all galaxies have 24$\mu$m and/or FUV detections ($N=47$) from which to derive internal extinctions. For these galaxies we adopt dust extinctions derived from $A_{\rm{H\alpha}}$ using the prescription of \citet{lee09b}. Using a sample of $\sim$300 nearby star-forming galaxies with robust spectroscopic Balmer decrement measurements (i.e., $f_{\rm{H\alpha}}/f_{\rm{H\beta}}$ flux ratios), \citet{lee09b} derived an empirical scaling relationship between $A_{\rm{H\alpha}}$ and $M_B$:

\begin{equation}
  A_{\rm{H\alpha}} =  
    \begin{cases}
    0.10 			   &  \textrm{if $M_B > -$14.5}\\
    1.971 + 0.323M_B + 0.0134M_B^2 &  \textrm{if $M_B \leq -$14.5}.
   \end{cases}
\end{equation}

\noindent We use this relationship to calculate $A_{\rm{H\alpha}}$ for the 47 galaxies with no 24$\mu$m and/or FUV detections and derive all other bandpass extinctions for these galaxies using the dust reddening curve of \citet{draine03}. \citet{lee09b} quotes a 20\% scatter for galaxies with $-14.7 > M_B > -18$ and only 10\% for galaxies fainter than $-14.7$. Since the majority ($N=36$) of these 47 galaxies are low-luminosity dwarfs ($M_B>-$14.7) and the remaining galaxies are fainter than $M_B > -18$, we do not expect these galaxies to add a significant amount of scatter to the trends presented in \S\ref{sec:results} and \S\ref{sec:discussion}. Quantities in this study which have been corrected for internal extinction are denoted with a subscript ``0" and represent an intrinsic measurement. The $A_{\rm{FUV}}$ values derived in this study (including those derived from $A_{\rm{H\alpha}}$) are presented in Table~\ref{tab:genprop}.

We have also compared our $A_{\rm{FUV}}$ determinations to another empirically derived method utilizing the TIR-to-FUV luminosity ratio ($L_{TIR}$/$L_{FUV}$) of \citet{buat05}. The TIR measurements are taken from the published values of \citet{dale09}. We find no significant change in the relationships presented in this study when using the $A_{\rm{FUV}}$ values derived from the prescription of \citet{hao11} and those of \citet{buat05}. Furthermore, despite several galaxies with significant dust extinction corrections ($A_{\rm{FUV}} > 0.5$ mag), the majority ($N=195$) of our galaxies exhibit small $A_{\rm{FUV}}$ values ($A_{\rm{FUV}} < 0.5$ mag). Consequently, all figures presented in \S\ref{sec:results} and \S\ref{sec:discussion} show similar overall structure with minor increased scatter when no dust corrections are applied. That is, our choice of dust extinction method and extinction curve does not significantly affect our results.

\subsection{SFR \label{sec:sfr}}
The SFR of galaxies can be estimated from many different luminosity tracers (e.g., H$\alpha$, FUV, etc.), where measured luminosities are transformed into SFRs via scaling prescriptions \citep[e.g.,][]{kennicutt98,murphy11}. 

H$\alpha$ nebular emission-based SFRs have shown inconsistencies with other tracers in low-mass, low-SFR activity galaxies \citep[$SFR<0.1~\rm{M_{\odot}yr^{-1}}$; e.g.,][]{meurer09,lee09b}. Subsequent studies have found that these inconsistencies may be attributed to stochastic star formation \citep[e.g.,][]{slug,fumagalli11,dasilva12,eldridge12,andrews13,hermanowicz13,dasilva14} and/or a bursty star formation history \citep[e.g.,][]{weisz12,barnes13}. These results make an H$\alpha$-based SFR unsuitable for our dwarf-dominated sample. 

Non-ionizing continuum FUV-based SFRs provide a more robust measure \citep{salim07,lee11,dasilva14}, but these photons are extinguished to a greater degree than other SFR indicators. However, we have IR data available from which to calculate this extinction (see previous section) and hence can provide an accurate dust-corrected SFR based on FUV fluxes. 

For these reasons we utilize an FUV-based SFR for the LVL sample. The FUV fluxes are taken from work of \citet{lee11} and we correct for internal extinction as discussed in \S\ref{sec:ext}. We use the prescription of \citet{murphy11} to convert FUV fluxes into a SFR where the uncertainties on the SFR include only the photometric uncertainties. The SFR prescriptions of \citet{murphy11} are an updated version of \citet{kennicutt98} using a Kroupa \citep{kroupa01} IMF. The calibration coefficient of \citet{kennicutt98} is a factor of $\sim$1.6 larger than the \citet{murphy11} coefficient \citep[see][]{calzetti07}.

\subsection{$M_{\star}$ \label{sec:mstar}}
In general, the stellar mass ($M_{\star}$) of a galaxy can be obtained through empirically or theoretically derived mass-to-light ratios ($\Upsilon_{\star}$). The empirical method compares the light of a galaxy at a particular wavelength to the kinematic mass using the baryonic Tully-Fisher relationship \citep[e.g.,][]{mcgaugh00,verheijen01,mcgaugh05}, and the theoretical method utilizes population synthesis models to derive relationships between stellar mass and observable galaxy luminosities \citep[e.g.,][]{BDJ01,bell03,portinari04,zibetti09,meidt14}. However, comparisons of model stellar mass-to-light ratios using different bandpasses do not always provide self-consistent results, and hence provide inconsistent stellar mass estimates \citep[e.g.,][]{mcgaugh14}. 


Ideally, knowledge of every star's mass and luminosity in a specific bandpass would provide an accurate $\Upsilon_{\star}$. However, the closest method to this is constructing star formation histories using individually resolved stars in a galaxy \citep[e.g.,][]{weisz08,mcquinn10,weisz11}. The study of \citet{eskew12} utilized the mass estimates from HST-derived star formation histories and compared it to the 3.6$\mu$m luminosities of many LMC regions. This study found a $\Upsilon_{\star}^{3.6\mu m}$ of 0.5 where the scatter in these regions translates into a 30\% uncertainty in stellar mass. Due to the emerging consistency of a nearly constant $\Upsilon_{\star}^{3.6\mu m} \sim 0.5$ using a variety of methods \citep[e.g.,][Barnes et al. 2014; submitted]{oh08,eskew12,mcgaugh14,meidt14}, we have determined our stellar masses via the prescription $\Upsilon_{\star}^{3.6\mu m}=0.5~$M$_{\odot}/$L$_{\odot,3.6\mu m}$, where M$_{\odot}/$L$_{ \odot,3.6\mu m}$ is the mass-to-light ratio in units of solar masses per the solar luminosity in the Spitzer 3.6$\mu m$ filter bandpass \citep[L$_{ \odot,3.6\mu m}=1.4\times10^{32}$~erg s$^{-1}$;][]{oh08}. The stellar masses derived here are presented in Table~\ref{tab:genprop}. There is likely some contamination of the 3.6$\mu m$ luminosity from PAH 3.3$\mu m$ emission, but the contribution is small for low-mass galaxies and is on the order of 10\% for the most massive galaxies in our sample \citep{meidt12a}.


\subsection{Metallicity \label{sec:met}}


We utilize the metallicity catalog of \citet{marble10} which compiled the metallicities for 129 LVL galaxies from the literature. These metallicities are derived from ``direct", ``strong-line", or a mix of both methods \citep[for details see][]{marble10}. The majority of the ``strong-line" metallicities are on similar calibration scales \citep[i.e., similar to][]{mcgaugh91} and we expect potential offsets of $<$0.3 dex between ``direct" and ``strong-line" values \citep{berg12}.

We have updated the \citet{marble10} catalog with 30 ``direct" oxygen abundances measured by \citet{berg12}. This update replaces previous measurements (both ``direct" and ``strong-line") with the ``direct" \citet{berg12} measurements derived from high signal-to-noise spectra. Furthermore, \citet{berg12} also measured the ``strong-line" oxygen abundances of several galaxies with no [OIII] 4363 line detections and provided four different ``strong-line" calibration values. We have updated the \citet{marble10} catalog with 12 ``strong-line" measurements from \citet{berg12} when there was no previous measurement or the previous measurement was a ``strong-line" calibration. To be consistent with the majority of the \citet{marble10} ``strong-line" calibrations, we used the theoretical \citet{mcgaugh91} calibrated metallicities provided by \citet{berg12}. 

In addition, we have also added 6 new oxygen abundances from \citet{moustakas10} where both an empirical \citep{pt05} and theoretical calibration measurements \citep{kk04} are available. We add the oxygen abundances derived from the \citet{kk04} calibration since this calibration is based upon the theoretical \citet{mcgaugh91} method. The final metallicity sample is composed of roughly half ``direct" and half ``strong-line" values which are tabulated in Table~\ref{tab:genprop}.

\subsection{PAH \label{sec:pah}}
For many galaxies, starlight strongly contributes to the 8$\mu$m bandpass flux which necessitates removal of the underlying stellar continuum. The stellar continuum-subtraction is carried out via the prescription of \citet{helou04}: 

\begin{equation}
f_{\nu}^{\rm dust}(8.0\mu m) = f_{\nu}^{\rm obs}(8.0\mu m) - \eta \times f_{\nu}^{\rm obs}(3.6\mu m),
\label{eqn:pahcorr}
\end{equation}

\noindent where the scale factor $\eta=0.232$ \citep[see also;][]{engelbracht08,ciesla14}, $f_{\nu}^{\rm dust}(8.0\mu m)$ is the stellar-free ``dust-only" emission that falls within the 8$\mu m$ band pass, and $f_{\nu}^{\rm obs}(8.0\mu m)$ and $f_{\nu}^{\rm obs}(3.6\mu m)$ are the observed 8$\mu m$ and 3.6$\mu m$ flux densities; all $f_{\nu}$ values are in units of Jy. The median stellar continuum correction is 43\% of the observed value. All 8$\mu m$ luminosities which have been corrected for the underlying stellar continuum in the rest of this paper will have a superscript ``dust" (e.g., $L^{\rm dust}_{8}$ is stellar continuum-subtracted 8$\mu m$ luminosity and $M^{\rm dust}_{8}$ is the stellar continuum-subtracted 8$\mu m$ magnitude). However, the 8$\mu m$ data points in the plotted SEDs of \S\ref{sec:seds} will not have their stellar continuum subtracted. 

We have also compared our stellar continuum-subtracted 8$\mu m$ luminosities ($L^{\rm dust}_{8}$) to those derived by \citet{marble10}. \citet{marble10} derived their own stellar continuum component similar to the method of \citet{engelbracht08}, but subtracted off the dust continuum as well. The corrected luminosities of \citet{marble10} are lower due to the additional dust continuum-subtraction, but the stellar continuum component is in good agreement with values used in this study. In addition, the figures presented in \S\ref{sec:results} and \S\ref{sec:discussion} show no noticeable change in overall structure when using our stellar-continuum subtraction and the stellar plus dust-continuum subtraction of \citet{marble10}.

\section{Results } \label{sec:results} 
In this section we present relationships between observable and physical properties. We also present the panchromatic (1500~$\rm{\AA}-24~\mu$m) SEDs of the LVL galaxies binned by various physical properties. These SEDs provide a graphical representation of how underlying physical properties can influence the observed properties of a galaxy. The fluxes presented in this section have been corrected for internal dust extinction and Milky Way foreground extinction.

\begin{figure}
  \begin{center}
  \includegraphics[scale=0.5]{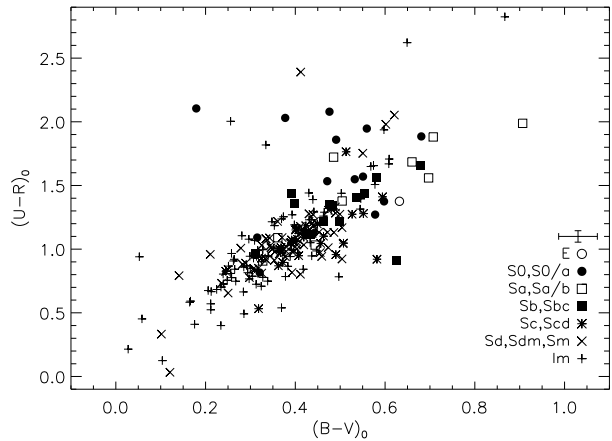}
  \caption{A color-color plot for the LVL composite $UBVR$ data where each filter magnitude has been corrected for internal dust extinction (see \S\ref{sec:ext}). The error bars above the legend represent the median errors of all data plotted. There is a positive relationship with significant overlap between different galaxy types. However, earlier-type galaxies tend to be redder in color.}
  \label{fig:color1}
  \end{center}
\end{figure}  

\begin{figure}
  \begin{center}
  \includegraphics[scale=0.5]{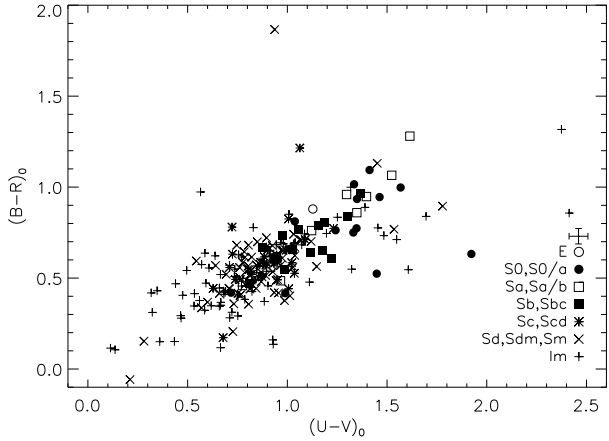}
  \caption{A color-color plot similar to that of Figure~\ref{fig:color1}. There is a positive relationship with significant overlap between different galaxy types. However, earlier-type galaxies tend to be redder.}
  \label{fig:color2}
  \end{center}
\end{figure}  

\subsection{Observable-Physical Relationships}\label{sec:optical}
Figures \ref{fig:color1} and \ref{fig:color2} show two optical color-color diagrams for the LVL data set. These relationships show a general positive trend between different colors. While galaxies of earlier morphological types tend to have redder colors, there is significant overlap in the color of the different types. 

\begin{figure}
  \begin{center}
  \includegraphics[scale=0.5]{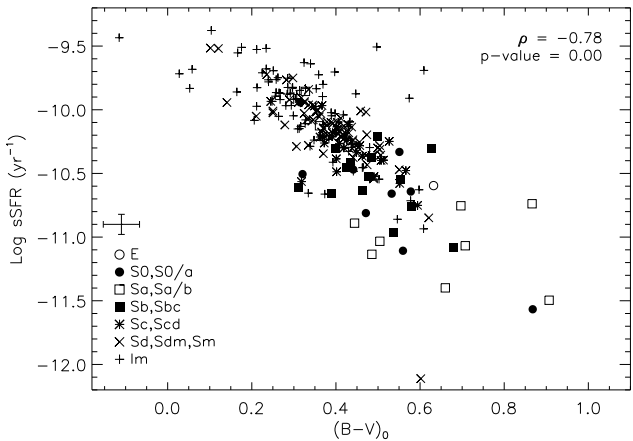}
  \caption{The sSFR versus $(B-V)_0$ color corrected for dust extinction. Bluer galaxies tend to have higher sSFR. Furthermore, later-type spirals and dwarfs tend to have higher sSFRs. The Spearman correlation coefficient ($\rho$) is provided and the corresponding p-value is indistinguishable from zero (i.e., $10^{-45}$).}
  \label{fig:BVsSFR}
  \end{center}
\end{figure}

Figure~\ref{fig:BVsSFR} shows a correlation between optical $(B-V)_0$ and specific star formation rate ($sSFR \equiv SFR/M_{\star}$), where optically blue galaxies tend to have higher sSFRs. Although there is some overlap between different galaxy types, lower-mass dwarf galaxies and later-type spirals tend to have bluer colors and higher sSFRs, while earlier-type galaxies tend to have redder colors and lower sSFRs. Also, the low scatter and moderately large Spearman correlation coefficient ($\rho=-0.78$) of this relationship indicates that extinction-corrected optical color could be used as a sSFR indicator in the absence of accurate SFR and stellar mass information \citep[see also,][]{schawinski14}.

\begin{figure}
  \begin{center}
  \includegraphics[scale=0.5]{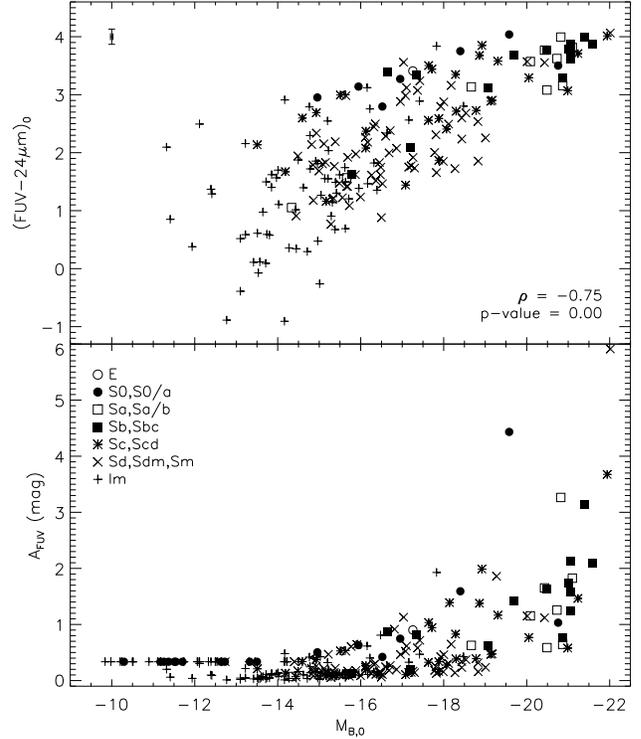}
  \caption{Top panel: the ($FUV-24\mu m$)$_0$ color versus absolute $B-$band magnitude; both corrected for dust extinction. Lower y-axis values represent less  warm dust emission and higher intrinsic FUV emission. The ($FUV-24\mu m$)$_0$ color is an indicator of the strength of young stellar emission with respect to  warm dust emission. The relationship shows that brighter galaxies inherently have higher warm dust to FUV ratios.  The Spearman correlation coefficient ($\rho$) is provided and the corresponding p-value is indistinguishable from zero. Bottom panel: the extinction in FUV due to dust in magnitudes versus the absolute $B-$band magnitude corrected for dust extinction. Brighter galaxies tend to be more opaque.}
  \label{fig:MbFUV24}
  \end{center}
\end{figure}  

Figure~\ref{fig:MbFUV24} shows a rough correlation between the dust-corrected absolute $B-$band magnitude ($M_{B,0}$) and the ($FUV-24\mu m)_0$ color with an increased scatter towards fainter (lower-mass) galaxies. The ($FUV-24\mu m)_0$ color represents the ratio of stellar UV photons to dust emission from small grains which are stochastically heated in regions of high intensity radiation fields; for brevity we refer to the $24\mu m$ emission as warm dust emission. The moderately large Spearman correlation coefficient ($\rho=-0.75$) of this trend suggests that brighter (i.e., more massive) star-forming galaxies tend to have a higher ratio of  warm dust-to-stellar UV emission. Consequently, brighter galaxies tend to have greater $A_{FUV}$ extinction corrections and hence more opaque \citep[see also,][]{buat99,calzetti01,sullivan01,buat02,moustakas06,weiner07,lee09b,momcheva13}. In addition, there is a systematic separation between galaxy types, where later-type spirals and dwarf galaxies are fainter and have reduced  warm dust emission compared to earlier-type galaxies.


\subsection{SEDs \label{sec:seds}}
\begin{figure*}
  \begin{center}
  \includegraphics[scale=0.6]{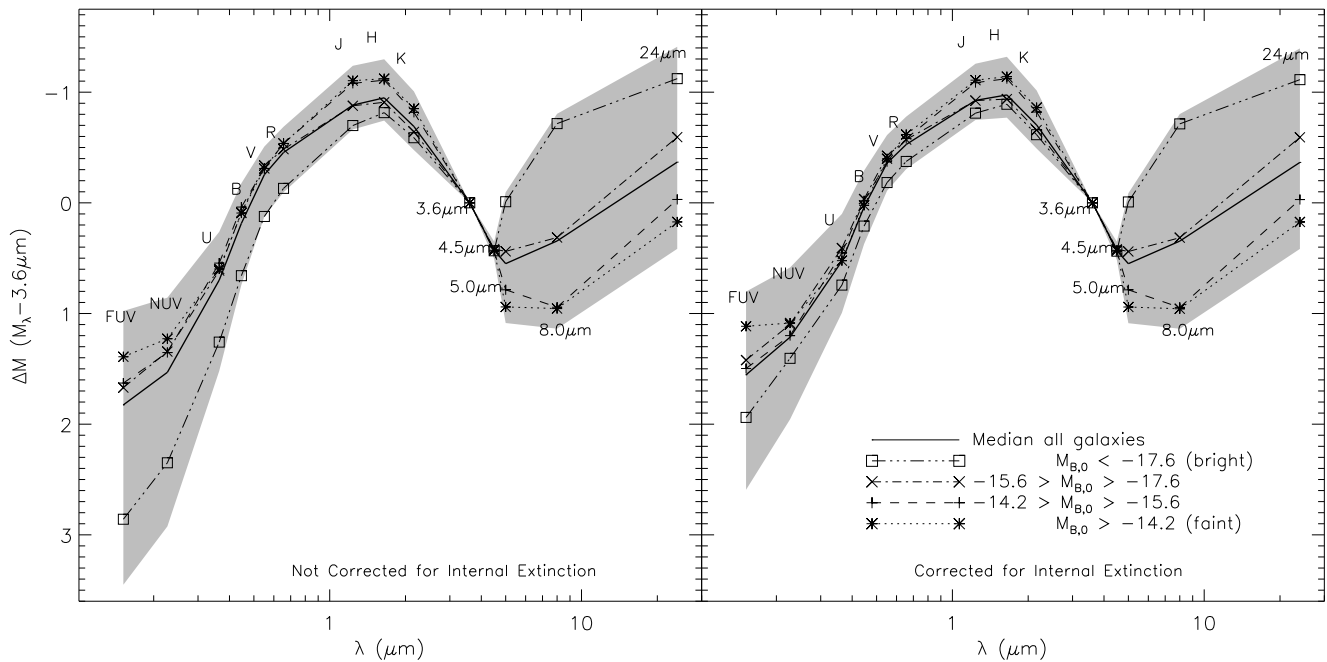}
  \caption{Spectral energy distribution (SED) of the LVL galaxy sample where all magnitudes have been normalized to the 3.6$\mu m$ measurement. The y-axis represents the difference in flux between each band pass and the 3.6$\mu m$ measurement. The galaxies in both panels have been binned by the absolute $B-$band ($M_{B,0}$) magnitude corrected for internal dust extinction. The left panel shows the LVL SEDs without dust correction while the right panel shows the dust-corrected SEDs. At each wavelength, the solid black lines represent the median value for all galaxies, the gray shaded regions represent the 16th and 84th percentile values, and the symbols represent the median flux of each bin.}
  \label{fig:sedMb}
   \end{center}
\end{figure*}

\begin{figure*}[u]
  \begin{center}
  \includegraphics[scale=0.6]{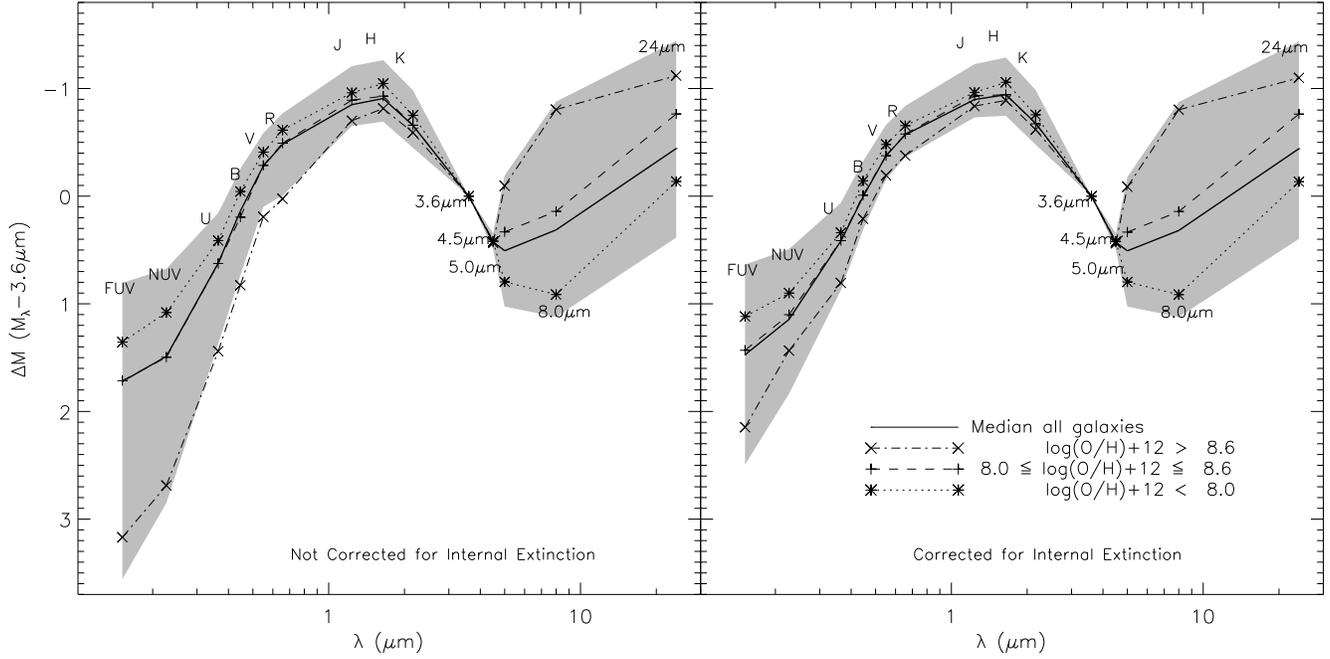}
  \caption{The LVL SEDs binned by 12+log(O/H) metallicity. The bins are defined by a separation of SMC (12+log(O/H)$\sim$8.0) and Solar (12+log(O/H)$\sim$8.6) metallicities instead of equal numbers of galaxies. However, the lowest, middle, and highest metallicity bins contain 67, 47, and 41 galaxies, respectively.}
  \label{fig:sedOH12}
  \end{center}
\end{figure*}

\begin{figure*}
  \begin{center}
  \includegraphics[scale=0.6]{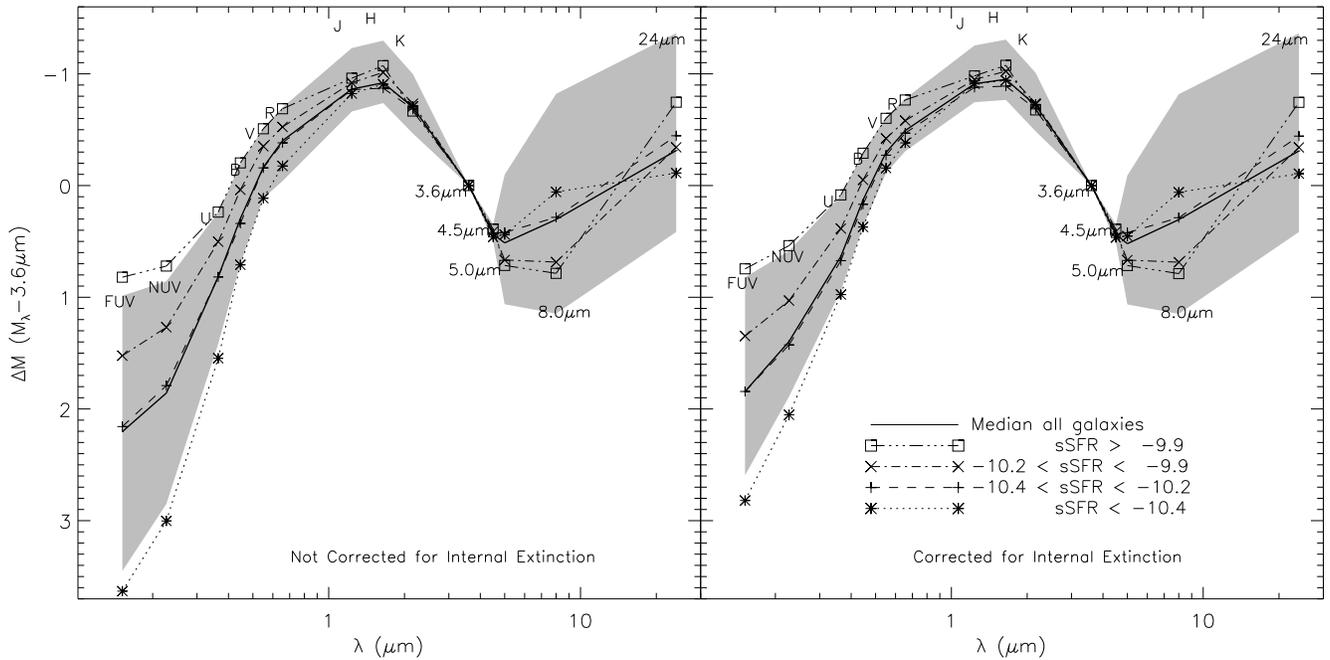}
  \caption{The LVL SEDs binned by the specific star formation rate (sSFR). The FUV-based SFR is corrected for extinction and $M_{\star}$ is based on the 3.6$\mu m$ measurement.}
  \label{fig:sedsSFR}
  \end{center}
\end{figure*}

This section presents the panchromatic ($FUV$ through $24\mu$m) SEDs in Figures \ref{fig:sedMb}--\ref{fig:sedsSFR} for the LVL galaxy sample binned by different galaxy properties. All fluxes are normalized to the 3.6$\mu$m measurement, where the 3.6$\mu$m flux represents emission from an older stellar population which constitutes the majority of a galaxy's stellar mass \citep{meidt12a}. Also, each figure has two panels comparing the SED shapes before (left panels) and after (right panels) correction for internal dust extinction. At each wavelength, the solid black lines and the gray shaded regions represent the median value and the 16th and 84th percentile values for all galaxies, respectively, and the symbols represent the median flux in each bin.

The bins in Figures \ref{fig:sedMb}--\ref{fig:sedsSFR} are defined so that each is populated with an equal number of galaxies to avoid any low number statistical biases except for the metallicity bins in Figure~\ref{fig:sedOH12} (see below). Furthermore, we base the bins upon fluxes, or properties derived from fluxes, which have been corrected for extinction due to dust (i.e., the right panels of Figures \ref{fig:sedMb}--\ref{fig:sedsSFR}). We have chosen to bin the galaxies in this manner since binning by non-dust-corrected fluxes (i.e., fluxes obscured by dust) could result in galaxies of similar intrinsic fluxes, or properties derived from similar intrinsic fluxes, to reside in separate bins. Consequently, the bins of both panels in Figures \ref{fig:sedMb}--\ref{fig:sedsSFR} will be comprised of the same galaxy population, but will exhibit different median fluxes since the fluxes in the left panel are not corrected for dust while those in the right panel are corrected for dust. These SEDs show how physical properties affect the observable properties of a galaxy sample before and after dust correction. 

Figure~\ref{fig:sedMb} shows LVL SEDs binned by ($M_{B,0}$) where brighter optical galaxies tend to have relatively brighter IR and fainter normalized UV fluxes. This systematic trend persists in the right panel of Figure~\ref{fig:sedMb}, where the fluxes have been corrected for internal dust extinction.


Figure~\ref{fig:sedOH12} presents the LVL SEDs binned by metallicity (12+log(O/H)), where each bin is separated by SMC (12+log(O/H)$\sim$8.0) and Solar (12+log(O/H)$\sim$8.6) metallicity values. These bins were chosen to represent commonly used values in the literature where each bin is populated with 67, 47, and 41 galaxies for the lowest, middle, and highest metallicity bins, respectively. Both panels show the same systematic trend, where metal-poor galaxies tend to be relatively brighter in the UV and fainter in the IR.

Figure~\ref{fig:sedsSFR} shows SEDs binned by sSFR where the normalized UV fluxes show a clear systematic trend in both panels: high sSFR galaxies are relatively bright in the UV and low sSFR galaxies are relatively faint in the UV. However, the normalized IR fluxes show a more complicated trend. The highest sSFR bin shows the lowest normalized $8\mu m$ fluxes and the highest normalized 24$\mu$m fluxes while the lowest sSFR bin has the highest normalized $8\mu m$ fluxes and the lowest normalized 24$\mu$m fluxes. 

\section{Discussion}\label{sec:discussion}
This section examines how dust properties are related to the observable and physical properties presented in Figures \ref{fig:sedMb}--\ref{fig:sedsSFR}. The latter part of this section utilizes the low-mass dominated LVL galaxy sample to test two ``main-sequence" galaxy relationships that have been established with higher mass galaxy samples. We find good agreement for the ``main-sequence" of star-forming galaxies, but find systematic deviations from the IR ``main-sequence."

\subsection{Dust}\label{sec:dust}
We discuss how dust affects the SEDs of Figures \ref{fig:sedMb}--\ref{fig:sedsSFR} in the next four sub-sections and present supporting relationships between dust, optical, and other galaxy properties. Stellar UV photons absorbed by dust are re-radiated at infrared wavelengths. The re-processing of UV photons by dust results in a reduced observed UV luminosity and enhanced 24$\mu$m luminosity (i.e., higher  warm dust emission). The ($FUV-24\mu$m)$_0$ color has been corrected for extinction due to dust and represents the relative strength between intrinsic UV and  warm dust emission.

A specific dust component examined later in this section is emission from PAH molecules. PAHs are made up of sheets of carbon and hydrogen and emit at many mid-IR wavelengths between 3 and $20\mu m$ due to stochastic heating from the interstellar radiation field \citep[e.g.,][and the many references therein]{leger84,allamandola89,DraineLi07}. These PAH features arise in most star-forming galaxies; however, many studies have found depressed PAH emission in low-metallicity, low-luminosity galaxies \citep[e.g.,][]{engelbracht05,jackson06,madden06,ohalloran06,wu06,draine07,smith07b,engelbracht08,galliano08,wu11,ciesla14}.

PAH molecules are vulnerable to photo-disassociation in harder radiation fields (i.e., high intensities of energetic UV photons). However, the formation of PAH molecules may also be suppressed due to low abundances of elements which make up PAHs. Essentially, the cause of low PAH emission in these galaxies is a question of nature (metallicity: lack of heavy elements) versus nurture (radiation field: destruction of PAHs). We find preliminary evidence of depressed PAH emission in our low-mass dominated galaxy sample in \S\ref{sec:sSFR}, but a more detailed investigation of the underlying cause is presented in \S\ref{sec:IRMS}. 

\subsubsection{$M_{\rm{B}}$}\label{sec:Mb}
The left panel of Figure~\ref{fig:sedMb} (not corrected for dust extinction) shows that galaxies binned by $M_{B,0}$ have a systematic trend in their SEDs: optically bright galaxies have diminished normalized UV and enhanced normalized IR fluxes when compared to optically faint galaxies. This trend suggests a relatively higher  warm dust emission for brighter galaxies \citep[see,][]{jackson06,rosenberg06,hong10}. Figure~\ref{fig:MbFUV24} in \S\ref{sec:optical} provides supporting evidence for this conclusion since it shows a positive correlation between $M_{B,0}$ and ($FUV-24\mu$m)$_0$, where brighter galaxies show greater warm dust emission per young stellar UV emission. This could be explained via metallicity arguments since brighter galaxies have higher metallicities \citep[e.g.,][]{tremonti04,hlee07,berg12}. Given similar radiation field intensities, metal-rich galaxies will exhibit higher warm dust emission due to a greater abundance of dust-forming elements (e.g., C, Fe, Si, Mg, etc.) compared to metal-poor galaxies (see \S\ref{sec:met2}). 

The persistence of the systematic UV trend in the right panel of Figure~\ref{fig:sedMb} after correcting for dust extinction shows that optically faint galaxies have brighter normalized intrinsic UV fluxes. This suggests that optically faint galaxies have bluer colors and more intense radiation fields compared to optically bright galaxies. This could also be the result of metallicity effects (see \S\ref{sec:met2}).



\subsubsection{Metallicity}\label{sec:met2}
\begin{figure}
  \begin{center}
  \includegraphics[scale=0.5]{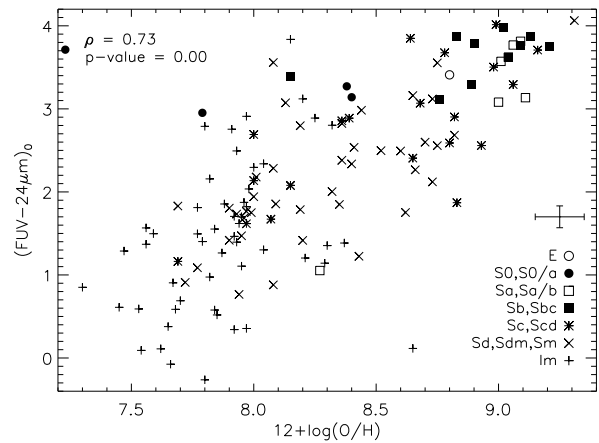}
  \caption{The ($FUV-24\mu m$)$_0$ color corrected for dust extinction versus metallicity (12+log(O/H)) where lower y-axis values represent less  warm dust emission and higher intrinsic FUV emission. There is a rough correlation between ($FUV-24\mu m$)$_0$ and metallicity. The correlation suggests that lower metallicity galaxies have relatively less  warm dust emission compared to higher metallicity galaxies. The Spearman correlation coefficient ($\rho$) is provided and the corresponding p-value is indistinguishable from zero.}
  \label{fig:OH12FUV24}
  \end{center}
\end{figure}

Figure~\ref{fig:sedOH12} shows that the highest metallicity bin has enhanced IR and reduced normalized UV fluxes similar to the trend found for brighter galaxies in \S\ref{sec:Mb}. This trend suggests a relative increase in  warm dust emission for higher metallicity galaxies. This could be the result of chemical enrichment where metal-rich galaxies will contain greater abundances of dust-forming elements and will exhibit higher  warm dust emission given similar radiation field intensities. Supporting evidence of this conclusion can be found in Figure~\ref{fig:OH12FUV24} where there exists a positive correlation between metallicity and ($FUV-24\mu$m)$_0$ color; redder ($FUV-24\mu$m)$_0$ colors (upward in the y-axis) correspond to higher oxygen abundances. This correlation is moderately strong with a Spearman coefficient of $\rho=0.73$. In addition, later-type spirals and dwarf (i.e., faint) galaxies tend to be metal-poor and exhibit diminished  warm dust emission compared to earlier-type galaxies.

The UV trend found in the uncorrected panel of Figure~\ref{fig:sedOH12} where low-metallicity galaxies tend to have higher normalized UV fluxes persists in the extinction corrected panel. This suggests that low metallicity galaxies have brighter normalized intrinsic UV fluxes compared to high metallicity galaxies; that is, low-metallicity galaxies tend to have more UV flux per unit stellar mass than higher metallicity galaxies. Higher metallicity galaxies will emit less in the UV, all other things equal, since metal-line blanketing reduces the photospheric UV luminosity of stars \citep[e.g.,][]{robert03}. Conversely, low metallicity stars will undergo less line-blanketing and thus produce relatively more UV photons (i.e., more intense radiation fields) in lower metallicity galaxies.

\subsubsection{sSFR}\label{sec:sSFR}
The SEDs of Figure~\ref{fig:sedsSFR} are binned by sSFR where there is a systematic trend in the UV: higher sSFR galaxies exhibit higher normalized UV fluxes. This trend persists in the dust-corrected panel suggesting that high sSFR galaxies have higher normalized intrinsic UV fluxes compared to low sSFR galaxies. Similar trends were found in previous sections inferring that faint, low-mass, low-metallicity galaxies have more intense radiation fields and higher sSFRs.

There is a complicated trend with sSFR in the IR where the highest sSFR bin shows high normalized 24$\mu$m fluxes but depressed normalized 8$\mu$m fluxes. In other words, the high sSFR bin shows a relative increase in warm dust emission, but a relative decrease in PAH emission compared to lower sSFR bins. This may be the result of more intense radiation fields destroying a larger portion of PAH molecules \citep[e.g., ][]{engelbracht05,madden06,wu06,engelbracht08,wu11}. 

\begin{figure}
  \begin{center}
  \includegraphics[scale=0.5]{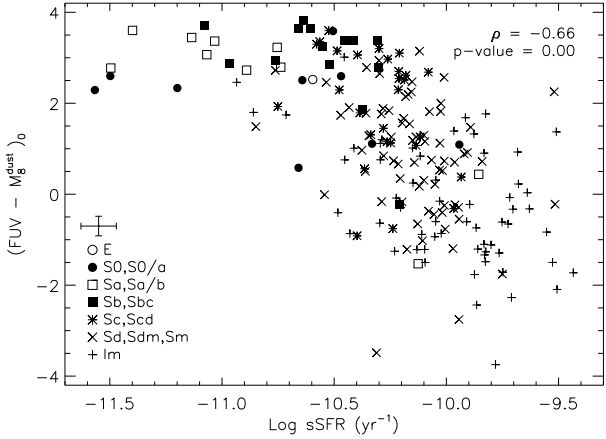}
  \caption{The ($FUV-M^{\rm dust}_{8})_0$ color corrected for dust extinction versus sSFR where lower y-axis values represent less PAH emission and higher intrinsic FUV emission. There is a clear correlation between ($FUV-M^{\rm dust}_{8})_0$ and sSFR. The correlation suggests that more intense radiation field environments (i.e., higher sSFRs) have reduced PAH emission. The Spearman correlation coefficient ($\rho$) is provided and the corresponding p-value is indistinguishable from zero.}
  \label{fig:FUV8sSFR}
  \end{center}
\end{figure}  

The possible destruction of PAHs in more intense radiation fields can also be seen in Figure~\ref{fig:FUV8sSFR} where there exists a relationship between sSFR and ($FUV-M^{\rm dust}_{8})_0$, where the ($FUV-M^{\rm dust}_{8})_0$ color is the ratio of FUV emission to stellar-subtracted 8$\mu$m dust emission. The moderately large Spearman correlation coefficient ($\rho=-0.66$) of the relationship in Figure~\ref{fig:FUV8sSFR} suggests that galaxies with bluer colors (i.e., more intense radiations fields) exhibit less PAH emission. Furthermore, later-type spirals and dwarf galaxies tend to have higher sSFRs compared to earlier-type galaxies. This trend supports the idea that low-mass, low-metallicity galaxies have more intense radiation fields and hence lower PAH emission. 

However, the effect of low metal abundance on PAH emission must also be considered since low-metallicities environments provide fewer starting materials (i.e, lower abundances of C) with which to form PAH molecules. The effect of PAH destruction and formation is further discussed in \S\ref{sec:IRMS}.

\subsection{``Main Sequence" of Star-Forming Galaxies}\label{sec:discMS}
In this section we test star formation relationships with the low-mass dominated LVL galaxy sample. These relationships have been established with higher-mass galaxy samples selected from both optical \citep{brinchmann04,salim07,elbaz07} and IR wavelengths \citep{elbaz11}.

\subsubsection{SFR vs. $M_{\star}$}\label{sec:MS}
\begin{figure}
  \begin{center}
  \includegraphics[scale=0.5]{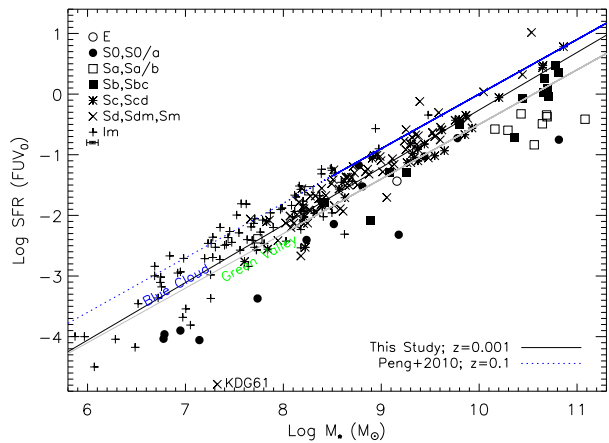}
  \caption{The ``main-sequence" of star-forming galaxies where there exists a tight correlation between SFR and stellar mass ($M_{\star}$). The relationship we find for the LVL galaxy sample is described by the equation: log$(SFR)= 0.95\pm0.004 \times $log$(M_{\star}) - 9.8\pm0.04$. We find good agreement with the relationship found by \citet{peng10} whose galaxy sample probes a similar galaxy population and redshift (median redshift of $\sim$0.1) as this study. The solid-dotted line transition represents lower end of the stellar mass range probed by \citet{peng10}.}
  \label{fig:MstarSFR}
  \end{center}
\end{figure}

Figure~\ref{fig:MstarSFR} presents the FUV-based SFR plotted versus $M_{\star}$ for the LVL sample. We compare our data to the low redshift ($z\sim0.1$) SDSS study of \citet{peng10} where the solid-dotted line transition in Figure~\ref{fig:MstarSFR} represents the lower end of the stellar mass range probed. The ``Main-sequence" found by \citet{peng10} agrees relatively well with our derived relationship: log$(SFR)= 0.95\pm0.004 \times $log$(M_{\star}) - 9.8\pm0.04$. There is an increasing small offset between our fit and \citet{peng10} toward lower stellar masses, but the difference is well within the scatter of our data (RMS=0.29~dex). It is interesting to note that the solid-blue line indicating the mass range of SDSS galaxies (log($M_{\star})\sim8.5~$M$_{\odot}$) studied by \citet{peng10} only extends half way through the mass range examined here. Since our low-mass ``Main-sequence" trend shows good agreement with the previous low-redshift study of \citet{peng10}, we conclude that the ``Main-sequence" of star-forming galaxies extends to the extreme environments of low-mass galaxies (log($M_{\star}) \sim 6~$M$_{\odot}$).


Based upon the DR7 data of \citet{peng10}, the study of \citet{schawinski14} characterized the SDSS ``green-valley" population where they found that the majority reside in an area $\sim$0.3$-$0.9 dex below the ``main-sequence" trend and that the SDSS ``main-sequence" galaxy contours extend to 0.5 dex below the ``main-sequence" trend (see Figure~5 of \citet{schawinski14}). The overlap between the SDSS ``main-sequence" and ``green-valley" galaxies makes it difficult to cleanly identify potential LVL ``green-valley" galaxies. However, a galaxy below the extent of the ``main-sequence" contours is unlikely to be a ``main-sequence" galaxy, hence it is more likely to be a ``green-valley" galaxy. We find a number LVL galaxies which fall 0.5 dex below the SDSS ``main-sequence" trend (i.e., below the \citet{peng10} line) who also exhibit redder optical colors than most of the LVL sample. These ``green-valley" LVL galaxies indicate that several local galaxies exhibit some degree of reduced star formation. 

In addition, there are several ``green-valley" LVL galaxies with stellar masses less than the SDSS mass range (log($M_{\star})<8.5~$M$_{\odot}$). These galaxies indicate that low-mass galaxies also show some degree of reduced star formation. The most notable of these is the dwarf galaxy KDG61 at log($M_{\star}) \sim 7.3$ and log($SFR) \sim$ $-$4.8. However, resolved star SFHs show that it is common for dwarf galaxies to undergo episodic star formation \citep[e.g.,][]{weisz08,mcquinn10,weisz11}, hence the low SFR of this galaxy could be due to a random occurrence of low activity and not due to a permanent evolution toward quenched star formation. This is supported by the recent SFH of KDG61 which shows a burst in SFR 1$-$2 Gyr ago and a present day SFR below the lifetime averaged SFR \citep{weisz11}. 




\subsubsection{IR ``Main Sequence"}\label{sec:IRMS}
\begin{figure}
  \begin{center}
  \includegraphics[scale=0.5]{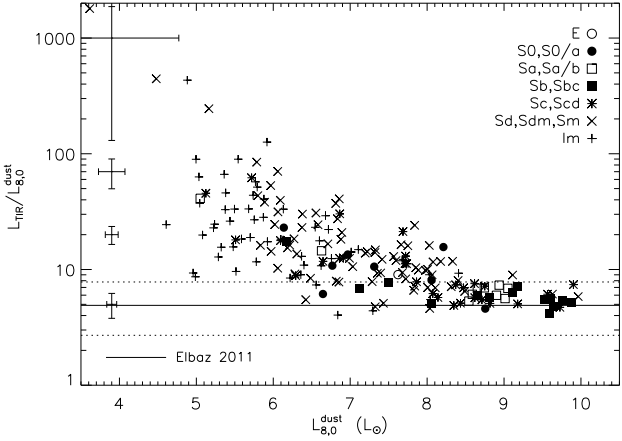}
  \caption{The IR ``main-sequence" relationship: total-IR (TIR)-to-8$\mu m$ luminosity ratio versus 8$\mu m$ luminosity. The PAH luminosity plotted has the underlying stellar continuum-subtracted. The error bars represent the median errors of evenly spaced TIR/PAH emission bins. Higher luminosity galaxies show good agreement with the relationship found by \citet{elbaz11} represented by the solid line and the corresponding 1$\sigma$ dotted lines. However, low luminosity galaxies show a systematic deviation from the IR ``main-sequence" suggesting a dearth in PAH emission.}
  \label{fig:TIR8v8}
  \end{center}
\end{figure}

The IR``main-sequence" relates the total-IR (TIR)-to-8$\mu m$ luminosity ratio ($L_{TIR}$/$L_{8}$) to the 8$\mu m$ luminosity ($L_{8}$) and has been established for bright galaxies \citep[$L_{8} > 10^8 $L$_{\odot}$;][]{elbaz11}. This sequence shows a constant $L_{TIR}$/$L_{8}$ ratio of $\sim$4.9$_{-2.2}^{+2.9}$ which seems to be independent of luminosity and redshift \citep{elbaz11}. However, \citet{lee13} showed deviations from this sequence at high luminosities ($L_{8} > 10^{11} $L$_{\odot}$) and \citet{dale09} found increased scatter towards fainter $M_{B}$ in the LVL sample. In order to directly test how low-mass galaxies behave on this relationship, we present the IR ``main-sequence" of the LVL galaxy sample in Figure~\ref{fig:TIR8v8} with axes defined the same as those of \citet{elbaz11}. 

Figure~\ref{fig:TIR8v8} reveals that galaxies at higher PAH luminosities ($L_{8} > 10^8 $L$_{\odot}$) show good agreement with the IR ``main-sequence." However, fainter (lower-mass) galaxies show a systematic deviation toward increased values by as much as a factor of 1000. This deviation suggests that the ``IR main sequence" does not extend to low-mass galaxies. Although \citet{elbaz11} did not correct for the underlying stellar continuum in their $L_8$ luminosities as we have done with our values, the systematic deviations are similar for observed (i.e., not corrected for stellar emission) LVL $L_8$ luminosities and exhibit increased $L_{TIR}$/$L_{8}$ values as large as 100. This systematic trend suggests either a relative increase in dust emission ($L_{TIR}$) or a relative decrease in PAH emission ($L_8$) in low-mass galaxies. 

Increased dust emission is unlikely in low-mass galaxies since Figure~\ref{fig:MbFUV24} shows that faint (low-mass) galaxies tend to be less opaque and Figure~\ref{fig:sedMb} shows that these galaxies have lower normalized warm dust emission at 24$\mu$m. Furthermore, the LVL study of \citet{dale09} showed a relatively flat 24$\mu$m-to-TIR flux ratio indicating that 24$\mu$m emission scales with TIR emission for star-forming galaxies, thus suggesting that low-mass galaxies with reduced 24 $\mu$m emission will have reduced TIR dust emission. If low-mass galaxies have reduced dust emission, then the increased IR ``main sequence" ratio ($L_{TIR}$/$L_{8}$) implies that the PAH emission is consequently reduced by an even greater amount.

\begin{figure}
  \begin{center}
  \includegraphics[scale=0.5]{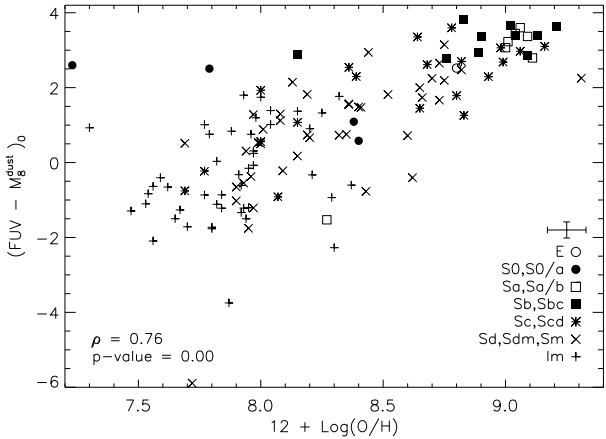}
  \caption{The ($FUV-M^{\rm dust}_{8})_0$ color versus metallicity. There is a correlation between ($FUV-M^{\rm dust}_{8})_0$ and metallicity. The correlation suggests that galaxies with lower metallicities have reduced PAH emission. The Spearman correlation coefficient ($\rho$) is provided and the corresponding p-value is indistinguishable from zero.}
  \label{fig:OH12FUV8}
  \end{center}
\end{figure}

Enhanced PAH suppression is plausible in low-mass galaxies and has been studied by many authors \citep[e.g.,][]{engelbracht05,jackson06,madden06,ohalloran06,wu06,draine07,smith07b,engelbracht08,galliano08,wu11,ciesla14}. There are two compounding factors that can conspire to depress the PAH emission in low-mass galaxies: destruction of PAHs in more intense radiation fields and suppressed PAH formation due to a dearth of starting material (i.e. C, Fe, Si, Mg, etc.) in low metallicity environments. The difficulty of disentangling these two effects can be visualized by comparing Figures~\ref{fig:FUV8sSFR} and \ref{fig:OH12FUV8}, where each figure plots the ($FUV-M^{\rm dust}_{8})_0$ color versus sSFR and metallicity, respectively. Both figures show trends with comparable scatter and correlation strengths where the Spearman coefficients are $\rho=-0.66$ and $\rho=0.76$ for Figures~\ref{fig:FUV8sSFR} and \ref{fig:OH12FUV8}, respectively.

\begin{figure*}
  \begin{center}
  \includegraphics[scale=0.5]{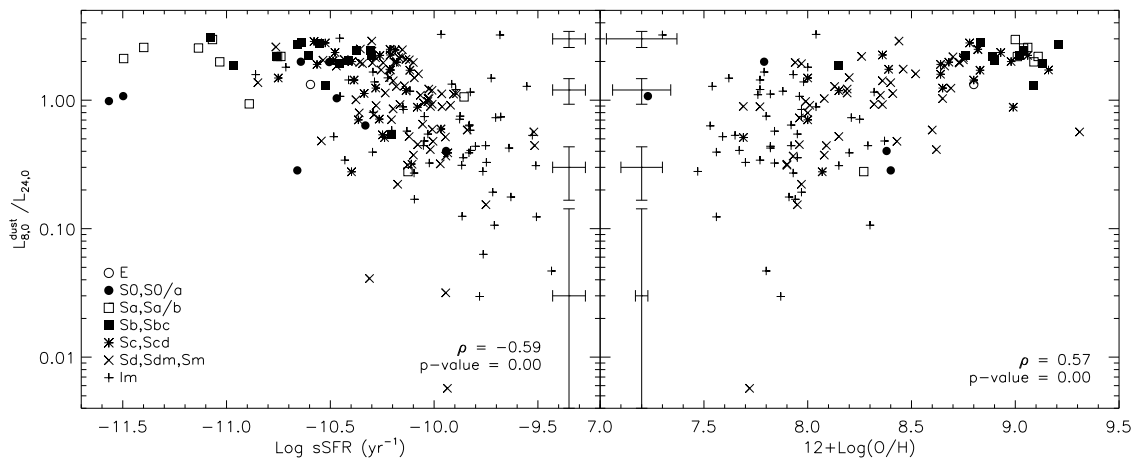}
  \caption{A comparison between the $L_{8}$/$L_{24}$ luminosity ratio and sSFR and metallicity in the left and right panel, respectively. The PAH measurements have been corrected for the underlying stellar continuum. Both relationships show similar trends suggesting that both sSFR and metallicity play similar roles in the reduction of PAH emission. The Spearman correlation coefficient ($\rho$) is provided and the corresponding p-value is indistinguishable from zero.}
  \label{fig:824vsSFROH12}
  \end{center}
\end{figure*}

Previous studies have also compared the $8\mu m$-to-24$\mu m$ luminosity ratio ($L_{8}$/$L_{24}$) to quantify reduced PAH emission in low metallicity environments \citep[e.g., ][]{engelbracht05,madden06}. Figure~\ref{fig:824vsSFROH12} shows $L_{8}$/$L_{24}$ plotted against both sSFR and metallicity in the left and right panels, respectively. Here we also find similar correlations with sSFR and metallicity, where both trends exhibit moderate scatter and moderately strong Spearman correlation coefficients ($\rho=-0.59$ and $\rho=0.57$ for sSFR and metallicity, respectively). 

\begin{figure*}
  \begin{center}
  \includegraphics[scale=0.5]{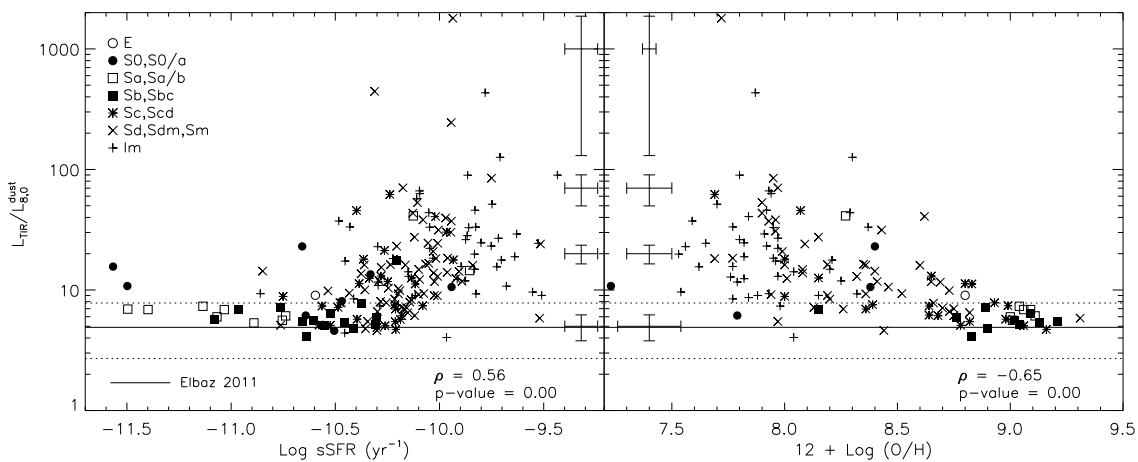}
  \caption{A comparison between the IR ``main sequence" $L_{TIR}$/$L_{8}$ luminosity ratio and sSFR and metallicity in the left and right panel, respectively. The PAH measurements have been corrected for the underlying stellar continuum. Both relationships show similar trends, but there exists a relatively clean upper sSFR envelope above log($sSFR$) $>$--10.2. The lack of data points above this value suggests a threshold in radiation field intensity above which PAH emission is diminished. The Spearman correlation coefficient ($\rho$) is provided and the corresponding p-value is indistinguishable from zero.}
  \label{fig:TIR8vsSFROH12}
  \end{center}
\end{figure*}

Further evidence of the effects of sSFR and metallicity on the depression of PAH emission in a galaxy can be seen by comparing the IR ``main sequence" $L_{TIR}$/$L_{8}$ ratio to sSFR (left panel) and metallicity (right panel) in Figure~\ref{fig:TIR8vsSFROH12}. There is a more pronounced increase in $L_{TIR}$/$L_{8}$ at log($sSFR$) greater than $\sim$ --10.8 compared to the gradual increase with metallicity. Also, it is interesting to note a nearly clean envelope in the sSFR panel of Figure~\ref{fig:TIR8vsSFROH12} to values of log($sSFR$) $>$ --10.2 to the right of which only two galaxies lie within the scatter of the IR main sequence relationship. This suggests a threshold of log($sSFR$) $\sim$ --10.2 above which the interstellar radiation field systematically destroys PAH molecules. Conversely, the nearly constant $L_{TIR}$/$L_{8}$ values toward lower log($sSFR$) $<$ --10.8 suggest either dust shielding from hard radiation fields or a lack of ubiquitous hard radiation fields to suppress PAH emission. However, the right panel of Figure~\ref{fig:TIR8vsSFROH12} shows a similar amplitude of increased $L_{TIR}$/$L_{8}$ scatter toward lower metallicities as seen with sSFR (left panel of Figure~\ref{fig:TIR8vsSFROH12}). This suggests that both metallicity and sSFR have comparable effects on the strength of PAH emission.

The effects of PAH destruction from more intense radiation fields (i.e., higher sSFRs) and suppressed PAH formation in low metallicity environments cannot be fully disentangled from these results. It is plausible that these two effects may not be distinguishable. \citet{sandstrom12} performed a spectroscopic study of the low-metallicity galaxy SMC and found that PAH molecules tend to be smaller on average compared to higher metallicity environments. Also, since small PAH molecules ($N_{\rm{Carbon}} < 50$) are more readily destroyed in typical background radiation fields \citep{allain96a,allain96b,lepage03,montillaud13,montillaud14}, it is likely that low metallicity environments facilitate the destruction of PAHs by preferentially forming smaller PAH molecules. 

It might be possible to determine which parameter (metallicity or radiation field hardness) has a greater effect on the emission strength of PAH molecules by controlling for a third parameter: the average size of the PAH molecules. A spectroscopic study of many star-forming regions within many galaxies would facilitate robust measurements of all three parameters via oxygen abundances for metallicity, flux ratios of different ionized species (e.g., Ne[III]/Ne[II]) for radiation field hardness, and flux ratios of different PAH emission lines which indicate the average PAH size \citep[e.g., $f_{7.7\mu m}/f_{11.3\mu m}$;][]{smith07b,sandstrom12}. The relative significance of metallicity and radiation field hardness might be disentangled via a similar study to the one performed here in regions with similar average sized PAH molecules \citep[see][]{haynes10}. However, this is beyond the scope of this study and is left for future projects.

A practical consequence of the analysis in this section suggests that the PAH luminosities of low-mass, low-metallicity galaxies are not a robust indicator of star formation rates \citep[see][]{calzetti07}. The systematic variation of the PAH emission combined with moderate scatter make the calibration of 8$\mu m$ luminosities as a SFR indicator highly uncertain.

\section{Summary}\label{sec:summary}
This study has constructed the panchromatic (UV-optical-IR) SEDs of the LVL galaxy sample which is dominated by low-mass dwarf and irregular galaxies. These galaxies show relationships between different optical colors (Figures \ref{fig:color1} and \ref{fig:color2}), optical color and sSFR (Figure~\ref{fig:BVsSFR}), and absolute $B-$band magnitudes and  warm dust emission normalized by young stellar emission (Figure~\ref{fig:MbFUV24}).

We have also examined the effects of dust on these SEDs (Figure~\ref{fig:sedMb}--\ref{fig:sedsSFR}) and found that optically bright galaxies tend to have relatively more warm dust emission and tend to have relatively low intrinsic normalized FUV fluxes compared to optically faint galaxies. The LVL SEDs have also revealed that low metallicity galaxies tend to have higher intrinsic UV and lower normalized IR fluxes compared to high metallicity galaxies (Figure~\ref{fig:OH12FUV24}). Furthermore, higher sSFR galaxies have higher normalized intrinsic UV fluxes compared to lower sSFR galaxies, but show complicated trends with normalized IR fluxes. The galaxies with high sSFRs show depressed normalized 8$\mu m$ fluxes (PAH emission) but high normalized 24$\mu m$ fluxes (warm dust emission) suggesting high sSFR environments modify or even destroy PAH molecules (Figure~\ref{fig:FUV8sSFR}). 

Testing previously established ``main-sequence" relationships of star-forming galaxies (SFR versus $M_{\star}$) with our sample of local Universe galaxies yielded good agreement with previous local Universe samples (Figure~\ref{fig:MstarSFR}). The LVL relationship is characterized by the equation: log$(SFR)= 0.95\pm0.004 \times $log$(M_{\star}) - 9.8\pm0.04$. The similar trends observed between our low-mass dominated galaxy sample and those of higher galaxy masses and similar redshifts indicate that the star-forming ``main sequence" extends to lower masses than previously probed. We also found several low-mass galaxies with lower SFRs at a given stellar mass indicating that these galaxies could be classified as ``green-valley" galaxies exhibiting reduced star formation. However, since dwarf galaxies tend to exhibit stochastic star formation activity in their star formation histories, the ``green-valley" dwarf galaxies may be exhibiting an episodic lull in activity and not a permanent transition to quenched star formation. 

We also test our low-mass dominated galaxy sample on the IR ``main-sequence" ($L_{TIR}$/$L_{8}$ versus $L_{8}$) which is established with the higher-mass galaxy sample of \citet{elbaz11}. We found systematic deviations toward higher $L_{TIR}$/$L_{8}$ at low luminosities ($L_{8} < 10^8 $L$_{\odot}$; Figure~\ref{fig:TIR8v8}) suggesting that the ``IR main sequence" does not extend to lower-mass galaxies. We attribute these deviations to reduced PAH emission instead of enhanced TIR emission since low-luminosity galaxies tend to have relatively reduced (warm) dust emission compared to higher-luminosity galaxies (Figure~\ref{fig:MbFUV24}). Reduced PAH emission can be caused by either destruction of PAH molecules in more intense radiation fields (i.e., higher sSFRs) or the suppression of PAH formation in low-metallicity environments due to an underabundance of elements which make up PAHs. 

We find comparable correlations in three relationships between PAH emission indicators and sSFR and metallicity (Figures \ref{fig:FUV8sSFR}, \ref{fig:OH12FUV8}, \ref{fig:824vsSFROH12}, \ref{fig:TIR8vsSFROH12}). Both ($FUV-M^{\rm dust}_{8})_0$ and $L_{8}$/$L_{24}$ luminosity ratios show similar trends with both sSFR and metallicity which suggests that these data cannot distinguish between the compounding effects of PAH destruction and suppressed PAH formation. However, the $L_{TIR}$/$L_{8}$ luminosity ratio shows a more pronounced increase with sSFR compared to a gradual increase with metallicity. This threshold is characterized by a log($sSFR$) value of $\sim$ --10.2 above which PAH molecules are systematically modified or destroyed. Finally, we speculate that the effects of these two factors may be linked since low-metallicity environments preferentially form smaller PAH molecules \citep{sandstrom12} which are more readily destroyed by interstellar radiation fields. 
  
\bibliographystyle{mn2e.bst}   
\bibliography{all.bib}

\end{document}